\documentclass[prl, twocolumn, showpacs]{revtex4}
\usepackage{graphicx}

\begin{document}

\title{A NEW OPTIMAL\ BOUND\ ON\ LOGARITHMIC\ SLOPE\ OF\ ELASTIC HADRON-HADRON SCATTERING}
\author{D. B. Ion$^{1,2)}$ and M. L. D. Ion$^{3)}$ \\
$^{1)}$TH Division, CERN, CH-1211 Geneva 23, Switzerland and\\
$^{2)}$NIPNE-HH, Bucharest P.O Box MG-6. Romania\\
$^{3)}$University of Bucharest, Department of Atomic and Nuclear Physics, Romania}

\begin{abstract}
In this paper we prove a new optimal bound on the logarithmic slope of the
elastic {\it slope b} {\it when: }$\sigma _{el}${\it \ and }$\frac{d\sigma }{%
d\Omega }(1)$ and $\frac{d\sigma }{d\Omega }(-1),$ {\it are known from
experimental data}. The results on the experimental tests of {\it this new
optimal bound }are presented in Sect. 3 for the principal meson-nucleon
elastic scatterings: ($\pi ^{\pm }P\rightarrow \pi ^{\pm }P$ and $K^{\pm
}P\rightarrow K^{\pm }P)$ at all available energies. Then we show that the
saturation of this optimal bound is observed with high accuracy practically at
all available energies in meson-nucleon scattering.
\end{abstract}

\maketitle

{\bf 1. Introduction}

Recently, in Ref. [1], by using {\it reproducing kernel Hilbert space }%
(RKHS\} methods [2-4], we described the quantum scattering of the spinless
particles by a {\it principle of minimum distance } {\it in the space of the
scattering quantum states (PMD-SQS). }Some preliminary experimental tests of
the {\it PMD-SQS, even }in the crude form [1], when the complications due to
the particle spins are neglected, showed that the actual experimental data
for the differential cross sections of all $PP,$ $\overline{P}P,$ $K^{\pm
}P, $ $\pi ^{\pm }P,$ scatterings at all energies higher than 2 GeV, can be
well systematized by {\it PMD-SQS }predictions{\it .} Moreover, connections
between the {\it optimal states [}1{\it ], the PMD-SQS in the space of
quantum states and }the {\it maximum entropy principle for the statistics of
the scattering channels }was also recently established by introducing {\it %
quantum scattering entropies } \cite{pl352}-\cite{prl83}.

The aim of this paper is to prove a new optimal bound on the logarithmic
 slope of the elastic hadron-hadron scattering by solving the following
optimization problem: {\it to find an lower bound on the
logarithmic slope b}
{\it when: }$\sigma _{el}${\it , }$\frac{d\sigma }{d\Omega }(+1)$ and $\frac{%
d\sigma }{d\Omega }(-1),$ {\it including spin effects, are given}. The results on the experimental
tests of {\it this new optimal bound }are presented for the principal
meson-nucleon elastic scatterings: ($\pi ^{\pm }P\rightarrow \pi ^{\pm }P$
and $K^{\pm }P\rightarrow K^{\pm }P)$ at all available energies. Then it was
shown that the saturation of this optimal bound is observed with high
accuracy practically at all available energies in meson-nucleon scattering.

{\bf 2}. {\bf Optimal helicity amplitudes for spin }$(0^{-}1/2^{+}%
\rightarrow 0^{-}1/2^{+})${\bf \ scatterings}

First we present some basic definitions and results for the optimal states
in the meson-nucleon scattering when the integrated elastic cross section $%
\sigma _{el}$ and differential cross sections $\frac{d\sigma }{d\Omega }(\pm
1)$ are known from experiments. Therefore, let $f_{++}(x)$ and $f_{+-}(x)$, $%
x\in [-1,1]$, be the scattering helicity amplitudes of the meson-nucleon
scattering process:
\begin{equation}
M(0^{-})+N(1/2^{+})\rightarrow M(0^{-})+N(1/2^{+})  \label{1}
\end{equation}
$x=\cos \theta ,\theta $ being the c.m. scattering angle. The formalizations
of the helicity amplitudes $f_{++}(x)$ and $f_{+-}(x)$ are chosen such that
the differential cross section $\frac{d\sigma }{d\Omega }(x)$ is given by
\begin{equation}  \label{2}
\frac{d\sigma }{d\Omega }(x)=\mid f_{++}(x)\mid ^2+\mid f_{+-}(x)\mid ^2
\end{equation}
Then, the elastic integrated cross section $\sigma _{el}$ is given by
\begin{equation}
\frac{\sigma _{el}}{2\pi}= \int_{-1}^{+1}\frac{d\sigma }{d\Omega
}(x)dx= \int_{-1}^{+1}[\mid f_{++}(x)\mid ^{2}+\mid f_{+-}(x)\mid
^{2}]dx  \label{3}
\end{equation}
Since we will work at fixed energy, the dependence of $\sigma _{el}$ and,$%
\frac{d\sigma }{d\Omega }(x)$ and of $f(x)$, on this variable was
suppressed. Hence, the helicities of incoming and outgoing nucleons are
denoted by $\mu $, $\mu ^{^{\prime }}$, and was written as (+),(-),
corresponding to $(\frac{1}{2})$ and $(-\frac{1}{2})$, respectively. In
terms of the partial waves amplitudes $f_{J+}$ and $f_{J-}$ we have
\begin{equation}
\left\{
\begin{tabular}{l}
$f_{++}(x)=\sum_{J=\frac{1}{2}}^{J_{\max }}(J+\frac{1}{2})(f_{J-}+f_{J+})d_{%
\frac{1}{2}\frac{1}{2}}^{J}(x)$ \\
$f_{+-}(x)=\sum_{J=\frac{1}{2}}^{J_{\max }}(J+\frac{1}{2})(f_{J-}-f_{J+})d_{-%
\frac{1}{2}\frac{1}{2}}^{J}(x)$%
\end{tabular}
\right\}  \label{4}
\end{equation}
where the d$_{\mu \nu }^{J}(x)$-rotation functions are given by
\begin{equation}
\left\{
\begin{tabular}{l}
$d_{\frac{1}{2}\frac{1}{2}}^{J}(x)=\frac{1}{l+1}\cdot \left[ \frac{1+x}{2}%
\right] ^{\frac{1}{2}}\left[ P_{l+1}^{,}(x)-P_{l}^{,}(x)\right] $ \\
$d_{-\frac{1}{2}\frac{1}{2}}^{J}(x)=\frac{1}{l+1}\cdot \left[ \frac{1-x}{2}%
\right] ^{\frac{1}{2}}\left[ P_{l+1}^{,}(x)+P_{l}^{,}(x)\right] $%
\end{tabular}
\right\}  \label{5}
\end{equation}
and prime indicates differentiation of Legendre polinomials $P_{l}(x)$ with
respect to x $\equiv \cos \theta $.
\begin{equation}
\frac{\sigma _{el}}{2\pi }=\sum (2J+1)\left[
|f_{J^{+}}|^{2}+[f_{J^{-}}|^{2}\right]  \label{6}
\end{equation}

Now, let us consider the optimization problem
\begin{equation}
\left\{
\begin{tabular}{l}
$\min \left[ \sum (2J+1)(|f_{J^{+}}|^{2}+|f_{J-}|^{2})\right] ,$ $\text{%
subject to: }$ \\
$\frac{d\sigma }{d\Omega }(+1)=fixed,$and $\frac{d\sigma }{d\Omega }%
(-1)=fixed$%
\end{tabular}
\right\}  \label{7}
\end{equation}
which will be solved by using Lagrange multiplier method [9] where
\begin{equation}
\left\{
\begin{tabular}{l}
$\pounds =\left[ \sum (2J+1)(|f_{J^{+}}|^{2}+|f_{J-}|^{2})\right] $ \\
$+\alpha \left[ \frac{d\sigma }{d\Omega }(+1)-|\sum
(J+1/2)(f_{J-}+f_{J+})|^{2}\right] $ \\
$+\beta \left[ \frac{d\sigma }{d\Omega }(-1)-|\sum
(J+1/2)(f_{J-}-f_{J+})|^{2}\right] $%
\end{tabular}
\right\}  \label{8}
\end{equation}

So, we prove that the solution of the problem (7)- (8) is as follows
\begin{equation}
\left\{
\begin{tabular}{l}
$f_{++}^{o}(x)=f_{++}(+1)\cdot \frac{K_{\frac{1}{2}\frac{1}{2}}(x,+1)}{K_{%
\frac{1}{2}\frac{1}{2}}(+1,+1)}$ \\
$f_{+-}^{o}(x)=f_{+-}(-1)\cdot \frac{K_{-\frac{1}{2}\frac{1}{2}}(x,-1)}{K_{-%
\frac{1}{2}\frac{1}{2}}(-1,-1)}$%
\end{tabular}
\right\}  \label{9}
\end{equation}
where the reproducing kernel functions are defined as
\begin{equation}
\left\{
\begin{tabular}{l}
$K_{\frac{1}{2}\frac{1}{2}}(x,y)=\sum_{\frac{1}{2}}^{J_{o}}(J+\frac{1}{2})d_{%
\frac{1}{2}\frac{1}{2}}^{J}(x)d_{\frac{1}{2}\frac{1}{2}}^{J}(y)$ \\
$K_{-\frac{1}{2}\frac{1}{2}}(x,y)=\sum_{\frac{1}{2}}^{J_{o}}(J+\frac{1}{2}%
)d_{-\frac{1}{2}\frac{1}{2}}^{J}(x)d_{-\frac{1}{2}\frac{1}{2}}^{J}(y)$ \\
\begin{tabular}{l}
$2K_{\frac{1}{2}\frac{1}{2}}(+1,+1)=(J_{o}+1)^{2}-1/4$ \\
$2K_{-\frac{1}{2}\frac{1}{2}}(-1,-1)=(J_{o}+1)^{2}-1/4$%
\end{tabular}
\\
$(J_{o}+1)^{2}-\frac{1}{4}=\frac{4\pi }{\sigma _{el}}\cdot \left[ \frac{%
d\sigma }{d\Omega }(1)+\frac{d\sigma }{d\Omega }(-1)\right] $%
\end{tabular}
\right\}   \label{10}
\end{equation}
Proof: Let us consider the complex partial amplitudes $f_{J^{\pm }}\equiv
r_{J^{\pm }}+ia_{J^{\pm }},$where $r_{J^{\pm }}$\ and $a_{J^{\pm }}$ are
real and imaginary parts, respectively. Then, Eq.(8) can be expressed
completely in terms of the variational variables $r_{J^{\pm }}$\ and $%
a_{J^{\pm }}$. Therefore, by calculating the first derivative we obtain
\begin{equation}
\left\{
\begin{tabular}{l}
$\frac{1}{(2J+1)}\frac{\partial \pounds }{\partial r_{J^{\pm }}}=r_{J^{\pm
}}-\alpha R^{++}(+1)\pm \beta R^{+-}(-1)=0$ \\
$\frac{1}{(2J+1)}\frac{\partial \pounds }{\partial a_{J^{\pm }}}=a_{J^{\pm
}}-\alpha A^{++}(+1)\pm \beta A^{+-}(-1)=0$%
\end{tabular}
\right\}   \label{11}
\end{equation}
where we have defined $f^{++}(x)\equiv R^{++}(x)+iA^{++}(x),$ and $%
f^{+-}(x)\equiv R^{+-}(x)+iA^{+-}(x)$, respectively, where
\begin{equation}
\begin{tabular}{l}
$R^{++}(+1)=\sum (J+\frac{1}{2})(r_{J^{+}}+r_{J^{-}})$ \\
$A^{++}(+1)=\sum (J+\frac{1}{2})(a_{J^{+}}+a_{J^{-}})$ \\
$R^{+-}(-1)=\sum (J+\frac{1}{2})(r_{J^{-}}-r_{J^{+}})$ \\
$A^{+-}(-1)=\sum (J+\frac{1}{2})(a_{J^{-}}-a_{J+})$%
\end{tabular}
\label{12}
\end{equation}
Therefore, from Eqs (11) we get
\begin{equation}
\left\{
\begin{tabular}{l}
$r_{J+}=\alpha R^{++}(+1)-\beta R^{+-}(-1)$ \\
$r_{J-}=\alpha R^{++}(+1)+\beta R^{+-}(-1)$ \\
$a_{J+}=\alpha A^{++}(+1)-\beta A^{+-}(-1)$ \\
$a_{J^{-}}=\alpha A^{++}(+1)+\beta A^{+-}(-1)$%
\end{tabular}
\right\}   \label{13}
\end{equation}
Then, using the definitions (2) and (3), we get
\begin{equation}
\alpha ^{-1}=\beta ^{-1}=(J_{o}+1)^{2}-1/4=\frac{4\pi }{\sigma _{el}}\left[
\frac{d\sigma }{d\Omega }(+1)+\frac{d\sigma }{d\Omega }(-1)\right]
\label{14}
\end{equation}
and, consequently we obtain that the optimal solution of the problem (7) can
be written in the form
\begin{equation}
\begin{tabular}{l}
$f_{++}^{o}(x)=\frac{2f_{++}(+1)}{(J_{o}+1)^{2}-1/4}\sum_{1/2}^{J_{o}}(J+%
\frac{1}{2})d_{\frac{1}{2}\frac{1}{2}}^{J}(x)d_{\frac{1}{2}\frac{1}{2}%
}^{J}(+1)$ \\
$f_{+-}^{o}(x)=\frac{2f_{+-}(-1)}{(J_{o}+1)^{2}-1/4}\sum_{\frac{1}{2}%
}^{J_{o}}(J+\frac{1}{2})d_{-\frac{1}{2}\frac{1}{2}}^{J}(x)d_{-\frac{1}{2}%
\frac{1}{2}}^{J}(-1)$%
\end{tabular}
\label{15}
\end{equation}
Now from Eqs. (14) and (15) we obtain the optimal solution (9) in which the
reproducing functions $K_{\frac{1}{2}\frac{1}{2}}(x,y)$ and $K_{-\frac{1}{2}%
\frac{1}{2}}(x,y)$ are defined by (10).

{\bf 3}. {\bf Optimal bound on logarithmic slope }

We recall the definition of the elastic slope b, and the relation
\begin{equation}
b\equiv \frac{d}{dt}\left[ \ln \frac{d\sigma }{dt}(s,t)\right] |_{t=0}=\frac{%
\overline{\lambda }^{2}}{2}\frac{d}{dx}\left[ \ln \frac{d\sigma }{d\Omega }%
(x)\right] |_{x=1}  \label{16}
\end{equation}
where transfer momentum is defined by : $t=-2q^{2}(1-x),$ $\overline{\lambda
}=1/q,$and $q$ is the c.m momentum.

Now, let us assume that $\sigma _{el}$, $\frac{d\sigma }{d\Omega }(+1),$ and
$\frac{d\sigma }{d\Omega }(-1)$ are known from the experimental data. Then,
taking into account the solution (9)-(10) of the optimization problem (7),
it is easy to prove that the elastic slope b defined by (16) must obey the
optimal inequality:
\begin{equation}
b\geq b_{o}\equiv \frac{\overline{\lambda }^{2}}{4}\left\{ \frac{4\pi }{%
\sigma _{el}}\cdot \left[ \frac{d\sigma }{d\Omega }(+1)+\frac{d\sigma }{%
d\Omega }(-1)\right] -1\right\}  \label{17}
\end{equation}

{\it Proof:} Indeed a proof of the optimal inequality (17) can be obtained
as singular solution of the following optimization problem
\begin{equation}
\begin{tabular}{l}
$\min \left\{ b\right\} ,\text{ subject to: }\sigma _{el}=\text{fixed, }$ \\
$\frac{d\sigma }{d\Omega }(+1)=\text{fixed},\frac{d\sigma }{d\Omega }(-1)=%
\text{fixed}$%
\end{tabular}
\label{18}
\end{equation}

So, the lower limit of the elastic slope b is just the elastic of the
differential cross section given by the result (9)-(10). Consequently, we
obtain that the optimal slope $b_{o}$ is given by
\begin{equation}
b_{o}=\overline{\lambda }^{2}\frac{d}{dx}\left[ \frac{K_{\frac{1}{2}\frac{1}{%
2}}(x,+1)}{K_{\frac{1}{2}\frac{1}{2}}(+1,+1)}\right] |_{x=1}=\frac{\overline{%
\lambda }^{2}}{4}\left\{ \left[ J_{o}(J_{o}+2)-\frac{1}{4}\right] \right\}
\label{19}
\end{equation}
Then, using the second part of (14) we obtain the inequality (17).

An important model independent result obtained Ref. [1], via the{\it \ }%
description of quantum{\it \ } scattering{\it \ }by the{\it \ principle of
minimum distance in space of states (PMD-SS)}, is the following {\it optimal
lower bound on logarithmic slope of the forward diffraction peak} in
hadron-hadron elastic scattering:
\begin{equation}
b \geq b_{o}\geq \frac{\overline{\lambda }^{2}}{4}\left[ \frac{4\pi }{%
\sigma _{el}}\frac{d\sigma }{d\Omega }(1)-1\right]   \label{(20)}
\end{equation}
In is important to remark, the optimal bound (17) improves in a more general and exact form not
only the unitarity bounds derived by MacDowell and Martin \cite{McDow}
for the logarithmic slope $b_{A}$ of absorptive
contribution $\frac{d\sigma _{A}}{d\Omega }(s,t)$ to the elastic
differential cross sections but also the unitarity lower bound derived in
Ref. [1] (see also Ref. \cite{scf}, \cite{rjf49}) for the slope $b$ of the
entire $\frac{d\sigma }{d\Omega }(s,t)$
differential cross section. Therefore, it would be important to make an experimental
detailed investigation of the saturation of this bond in the hadron-hadron scattering,
especially in the low energy region.

{\bf 4.\ Experimental tests of the bound (17)}

A comparison of the experimental elastic slopes b with the {\it optimal slope%
} $b_{o}(17)$ is presented in Figs. 1 for ($\pi ^{\pm }P$ and $K^{\pm }P)$-%
{\it scatterings}: The values of the $\chi
^{2}=\sum_{j}(b_{j}-b_{oj})^{2}/(\epsilon _{bj}^{2}+\epsilon _{b_{o}j}^{2})$%
, (where $\epsilon _{bj}$ and $\epsilon _{b_{o}j}$ are the experimental
errors corresponding to $b$ and $b_{o}$ respectively)\ are used for the
estimation of departure from the {\it optimal PMD-SS-slope} $b_{o}$, and
then, we obtain the statistical parameters presented in Table 1. For $\pi
^{\pm }P$-{\it scattering} the experimental data on $b$, $\frac{d\sigma }{%
d\Omega }(+1),$ $\frac{d\sigma }{d\Omega }(-1),$and $\sigma _{el}$ , for the
laboratory momenta in the interval $0.2$ GeV$\leq p_{LAB}\leq 10$ $GeV$ are
calculated directly from the {\it phase shifts analysis} (PSA) of Hohler et
al. \cite{hohler}. To these data we added some values of $b$ from the linear fit of
Lasinski et al. \cite{lasinski} and also from the original fit of authors quoted in some references
in \cite{rjf49}.
Unfortunately, the values of $b_{o}$ corresponding to the Lasinski's data
\cite{lasinski} was impossible to be calculated since the values of $\frac{d\sigma }{%
d\Omega }(1)$ from their original fit are not given. For $K^{\pm }P-${\it %
scatterings }the experimental data on $b$, $\frac{d\sigma }{d\Omega }(+1),$ $%
\frac{d\sigma }{d\Omega }(-1)$ and $\sigma _{el}$ , in the case of $K^{-}P$,
are calculated from the {\it experimental }(PSA) solutions of Arndt et al.
\cite{arndt}. To these data we added those collected from the original fit of data
from references of \cite{rjf49} which the approximation $\frac{d\sigma }{%
d\Omega }(-1)=0$. For $K^{+}P$-scattering,
we added some values of $b$ from the linear fit of Lasinski
et al. \cite{lasinski} and also those pairs $(b,b_{o})$ calculated directly from the
{\it experimental }(PSA) solutions of Arndt et al. \cite{arndt}. All these results
can be compared with those presented in Ref. \cite{rjf49}.

{\bf 5}. {\bf Summary and Conclusion}

The main results and conclusions obtained in this paper can be summarized as
follows:

(i) In this paper we proved the optimal bound (17) as the singular solution (%
$\lambda _{0}=0)$ of the optimization problem {\it to find a lower bound on
the logarithmic slope b} {\it with the constraints imposed when }$\sigma
_{el}${\it \ and }$\frac{d\sigma }{d\Omega }(+1)$ and{\it \ }$\frac{d\sigma
}{d\Omega }(-1)$ {\it are fixed from experimental data. }This result is
similar with that obtained recently in Refs. \cite{1}, \cite{rjf49}  for the problem{\it \
to find an upper} {\it bound for the scattering entropies when} $\sigma _{el}
${\it \ and }$\frac{d\sigma }{d\Omega }(+1)${\it \ are fixed. }

(ii) We find that the optimal bound (17) is verified experimentally with high accuracy
at all available energies
for all the principal meson-nucleon scatterings.

(iii). From mathematical point of view, the {\it PMD-SQS}-{\it optimal states
}(9)-(10){\it , }are functions of {\it minimum constrained} {\it norm} and
consequently can be completely described by{\it \ reproducing kernel
functions (}see also Ref. [1,3-4]. {\it \ }So, with{\it \ }this respect the
{\it PMD}-SQS-{\it optimal states} from the {\it reproducing kernel Hilbert
space (RKHS) }of the scattering amplitudes are analogous to the {\it coherent%
} {\it states} from the RKHS of the {\it wave functions.}

(iv) The{\it \ PMD-SQS-optimal state} (9)-(10)  have not only the property
that is {\it the most forward-peaked quantum state }but also possesses many
other peculiar properties such as maximum Tsallis-like entropies, as well as
the scaling and the s-channel helicity conservation properties, etc.,  that
make it a good candidate for the description of the quantum scattering.via
an optimum principle. In fact the{\it \ }validity{\it \ of the principle of
least distance in space of states in hadron-hadron scattering}  is already
well illustrated in Fig. 1 and Table 1.

All these important properties of the optimal helicity amplitudes
(9)-(10) will be discussed in more detail in a forthcoming paper.

\begin{table*}
\caption{$\stackrel{\_}{\chi ^{2}}-$statistical parameters of
the principal hadron-hadron scattering. In these estimations for P$%
_{LAB}\leq 2$ GeV/c the errors $\epsilon _{b}^{PSA}(\pi ^{\pm }P)=0.1$ $%
b^{PSA}$ and $\epsilon _{b_{o}}^{PSA}(K^{\pm }P)=0.1b^{PSA}$ are taken into
account while for the errors to the optimal slopes b$_{o}$ calculated from
phase shifts analysis and \cite{rjf49}.}
\begin{tabular}{||l|l|l|l|l||}
\hline\hline
&  & For P$_{LAB}\geq 2GeV/c$ &  & For all P$_{LAB}\geq 0.2GeV/c$ \\
\hline\hline Statistical parameters & N$_{p}$ & $\quad \,\quad
\quad \chi ^{2}/$n$_{dof}$ & N$_{p}$ & $\quad \quad \chi
^{2}/$n$_{dof}$ \\ \hline\hline $\quad \pi ^{+}P\rightarrow \pi
^{+}P$ & 28 & \quad $\quad \quad $1.02 & 90 & $\quad \quad $3.37\\ \hline
$\quad \pi ^{-}P\rightarrow \pi ^{-}P$ & 31 & \quad $\quad
\quad $0.92 & 93 & \quad $\quad $8.00 \\ \hline
$\quad K^{+}P\rightarrow K^{+}P$ & 37 & \quad $\quad \quad $1.15 & 73 & $%
\quad \quad $1.91 \\ \hline
$\quad K^{-}P\rightarrow K^{-}P$ & 37
& \quad $\quad \quad $1.52 & 73 & $\quad \quad $7.84 \\
\hline
$\quad PP\rightarrow PP$ & 29 &
\quad $\quad \quad $5.01 & 32 & $\quad \quad $5.06 \\ \hline
$\quad \bar{P}P\rightarrow \bar{P}P$ & 27 & \quad $\quad \quad $0.56 & 45 & $%
\quad \quad $1.86 \\ \hline \hline
\end{tabular}
\end{table*}

\begin{figure*}
\includegraphics[width=12cm]{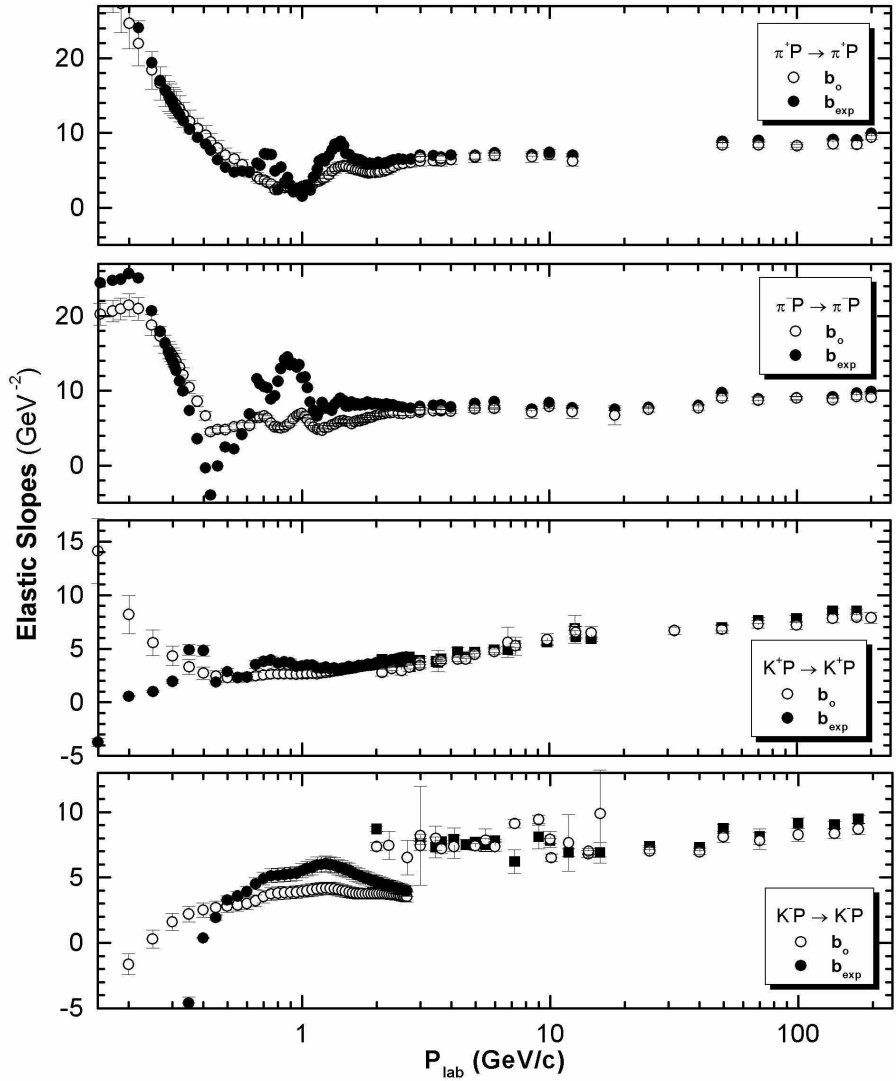}
\caption{The experimental values (black circles) of the
logarithmic slope b for the principal meson-nucleon scatterings
are compared with the {\it optimal PMD-SQS}-{\it predictions}
$b_{o}$ (white circles). The experimental data for $b${\bf ,
}$\frac{d\sigma }{d\Omega }(+1)$ and $\sigma _{el}$, are taken
from Refs. \cite{hohler}-\cite{lasinski}. (see the text).}
\end{figure*}

\end{document}